\documentclass{article}
\usepackage{spconf,amsmath,graphicx}
\usepackage{booktabs}
\usepackage{xeCJK}

\usepackage{tikz}
\usetikzlibrary{arrows.meta,positioning,decorations.pathreplacing}

\title{Boundary-Free Contextual Biasing: Depth-Adaptive Gating and Reading-Space Matching for Unsegmented Languages}
\name{Muhammad Huzaifah, Yu Pan, Zachary Yeo, Ningjie Bai, Guangzhao Yang}  %\thanks{Thanks to XYZ agency for funding.}
\address{R\&D Team, Recho Inc., Tokyo, Japan}
\begin{document}
\ninept
\maketitle
\begin{abstract}
Contextual biasing supplies an ASR system with a list of expected words at inference time, but existing methods rely on word boundaries that Japanese and Chinese do not provide. We present a boundary-free biasing decoder for frozen public CTC models, built on a character-level Aho–Corasick automaton, with no training and no second pass. Two evidence-based mechanisms replace the boundary: a depth-adaptive gate that sets how hard to push from match depth, and reading-space matching for when the audio is right but the characters are wrong. On Aishell-1 NE's hard R1 subset we reach 66.5\% recall, above the trained CLAS baseline (64\%), transferring to WenetSpeech and to a second architecture without retuning. We release the first open Japanese contextual-biasing benchmark, where biasing lifts rare-word recall by 25 points at precision above 97\%, and still by 19 and 22 points against 1,000-word lists.
\end{abstract}
\section{Introduction}
\label{sec:intro}

Even strong automatic speech recognition (ASR) models reliably miss the vocabulary their users care most about -- names, product terms, domain jargon -- rare in training data yet often carrying the utterance's meaning. \textit{Contextual biasing} addresses this with auxiliary information. A list of likely words (\textit{hotwords}) is supplied at inference and the decoder steered toward them. In production, models are frozen and lists can change per request, so steering must happen at decode time. A tight latency budget further rules out beam search and second-pass rescoring, leaving a single greedy pass over the CTC \cite{ctc} frame-synchronous token stream.

Any biasing decoder must decide, at every frame, where a listed word can start, whether the hypothesis is inside one and how far, and in what representation the match is scored. Segmented languages answer all three with one signal, the word boundary plus a stable written form, and most biasing literature is built on it \cite{aleksic15_interspeech,zhao19d_interspeech,sun2021tree,context_trie,andrusenko2025turbobias}. Unsegmented languages like Japanese (JA) and Chinese (ZH) answer none. First, no boundary token exists in the CTC stream, so every character position is a potential word entry. Second, no unit above the character is stable. The same string tokenizes context-dependently, so a listed word has no fixed token sequence, and CJK homophony means the model can decode the audio correctly and still emit the wrong characters (e.g. 照会→紹介). The stable object is the reading, which straddles token boundaries structurally. Matching must therefore run below the token, on characters at the surface and on morae or syllables in reading space. This buys no accuracy on its own (they tie in our experiments), but it makes the boundary's two lost roles recoverable. How hard to push follows from per-character match depth, which our gate switches on; what the push anchors to is the listed surface by default and, where homophony blocks it, per-mora or per-syllable reading state. Neither quantity exists at the word or token level.

No prior system exposes either quantity. Trained biasing attaches a module to the model (CLAS \cite{pundak2018deep}, TCPGen \cite{sun2021tree}, CB-Conformer \cite{xu2023cb}, SeACo-Paraformer \cite{shi2024seaco}); the ZH systems score each hotword as one learned embedding, leaving no incremental match state, at the cost of training data, model surgery and a fixed list distribution. Likewise phonetic biasing exists only at training time \cite{bruguier2019phoebe,chen2019joint} or as prompting \cite{peng24b_interspeech}; none runs a reading-space automaton on an unmodified checkpoint. Decode-time methods bypass training but anchor entries on boundaries and retune per list at scale \cite{aleksic15_interspeech,zhao19d_interspeech,context_trie,andrusenko2025turbobias}. For JA, WCTC-Biasing \cite{nakagome25_interspeech} is retraining-free but needs model-internal access, and its lists are error-mined from evaluation sets and unreleased. Open biasing sets exist for EN \cite{context_trie} and ZH \cite{xu2023cb,shi2024seaco,wang2025contextasr}, but not JA.

Our contributions are as follows: (1) A depth-adaptive biasing gate whose two extremes, restricting bias to the model's top-K candidates and letting it re-choose without a cap, are endpoints of one family switched by match-evidence depth in per-character information units. (2) Decode-time reading-space biasing, where the automaton advances on dictionary readings while bias promotes the listed surface, fused with the surface automaton by a correlation-discount rule. (3) The first open Japanese contextual-biasing benchmark, from JSUT \cite{sonobe2017jsut} and Common Voice JA \cite{ardila2020common}, stratified by hotword length and script (full construction methodology in a companion paper). (4) A cross-regime analysis on two architectures \cite{an2024funaudiollm,peng24b_interspeech} and two languages, characterizing failure modes on three axes: the threshold on script, the weight on list scale, and the cap on tokenizer granularity, and that one adaptive setting per language is robust across all of them.

%showing why unbounded gates are fragile on three independent axes and bounded gates are not.

% Our contributions are as follows. (1)~A depth-adaptive biasing gate whose two published extremes -- restricting bias to the model's top-K candidates, and letting it re-choose without a cap -- are endpoints of one family, switched by match-evidence depth in per-character information units. At the parents' own weight, one setting matches the best tuned parent within 1pp across scripts and list sizes without collapsing; weight, not gate, is the axis needing per-scale tuning. (2)~Decode-time reading-space biasing, where hotwords become mora or syllable paths, tokens advance by dictionary readings, and bias promotes the corresponding character surface, fused with the surface automaton in a single pass by a correlation-discount rule. (3)~The first open Japanese contextual-biasing benchmark, from JSUT [18] and Common Voice JA [19], stratified by hotword length and script, the regime variables our analysis identifies (construction methodology in a companion paper). (4)~A cross-regime analysis on two architectures and two languages showing why unbounded gates are fragile on three independent axes and bounded gates are not.

\section{Method}
\label{sec:method}

\subsection{Boundary-free surface biasing}
Hotwords are compiled into a character-level Aho–Corasick automaton \cite{aho-corasick}, a trie over the hotword set plus failure links to the longest matched suffix that is also a hotword prefix (Fig.~\ref{fig:automaton}a). It reports every occurrence in one left-to-right pass at a per-frame cost independent of list size. Because decoding emits tokens rather than characters, the automaton is precomputed into two token-level tables giving, for each state and token, the bias accrued by walking that token's surface and the state it reaches (Fig.~\ref{fig:automaton}b); a multi-character piece therefore crosses several edges in one emission. Decoding stays ordinary greedy CTC. At each frame the bias row $b(q_t,\cdot)$ is added to the log-probabilities $\ell_t$ and the emitted token $v_t$ advances the state, costing one row lookup, one transition and one integer register per stream, batched and streaming-compatible. Writing $d(q)$ for the depth of state $q$, its distance from $q_0$ and hence the number of characters matched so far, a state carries potential $\psi(d)=w\sum_{i=1}^{d}(2+\ln i)$, and a token moving the automaton from $q$ to $q'$ accrues $\psi(d(q'))-\psi(d(q))$ plus a lump on completing a hotword, as in \cite{andrusenko2025turbobias} but over characters. Credit therefore grows with match depth, the single weight $w$ scales the whole schedule, and negative deltas are clamped so abandoning a partial match is free. The failure links make every character position a possible entry, the boundary-free property, and let partial matches collapse without corrupting the hypothesis.

%Match scoring gives per-character advancement credit plus a completion lump, as in \cite{andrusenko2025turbobias} but over characters. 

%scaled by a single global weight $w$ that sets the bias magnitude, with negative deltas clamped so abandoning a partial match is free. 
% It reports every occurrence in one left-to-right pass at a cost independent of list size. 

% Match scoring uses potential-style credit, as in [13] but over characters, giving per-character advancement plus a completion lump, with negative deltas clamped so abandoning a partial match is free (charging for it was measured and rejected, §5)

\subsection{Depth-adaptive gating}

The gate replaces the boundary's control role, deciding how hard to push from the evidence the partial match carries. Two mechanisms bound the design space. The first, \textit{masked biasing}, is ours. Bias is added only among the model's top-K tokens before the argmax; making explicit a restriction beam search obtains from pruning \cite{zhao19d_interspeech}, it is entry-safe but unable to rescue a continuation ranked below K. The second, \textit{two-stage} (following TurboBias's greedy semantics \cite{andrusenko2025turbobias}, source-verified), re-chooses without a cap among candidates other than blank and the previous token; powerful far into a match but catastrophic at entries, where dense unmarked candidates let strong bias hijack ordinary text. Each fails outside its band (§4.3). A boundary supplies that distinction at zero depth, and \cite{zhao19d_interspeech} obtains it from word prefixes; without either, we infer it from match evidence.

%Beam search obtains this restriction accidentally from pruning, a defect [5] removes by biasing before pruning. Greedy decoding has no beam, so we impose it deliberately as the safety mechanism. It is entry-safe but capped far into a match, unable to rescue a continuation ranked below K. 

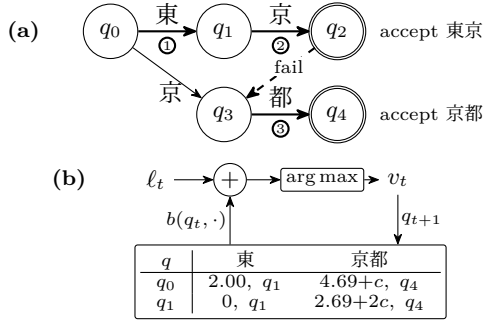
\begin{figure}[t]
\centering

%% ================= (a) character automaton =================
\begin{tikzpicture}[
  >={Stealth[round]}, font=\small,
  st/.style={circle,draw,minimum size=7.2mm,inner sep=0pt},
  ac/.style={st,double,double distance=0.5pt}]

  \node[font=\footnotesize\bfseries] at (-1.15,0.0) {(a)};

  \node[st] (q0) at (0,0)      {$q_0$};
  \node[st] (q1) at (1.50,0)   {$q_1$};
  \node[ac] (q2) at (3.00,0)   {$q_2$};
  \node[st] (q3) at (1.50,-1.1){$q_3$};
  \node[ac] (q4) at (3.00,-1.1){$q_4$};

  % edges on the 東京都 walk are drawn thick and numbered by input position
  \tikzset{stp/.style={circle,draw,inner sep=0.4pt,font=\tiny,fill=white}}
  \draw[->,thick] (q0) -- node[above=-1pt]{東} node[below=2pt,stp]{1} (q1);
  \draw[->,thick] (q1) -- node[above=-1pt]{京} node[below=2pt,stp]{2} (q2);
  \draw[->]       (q0) -- node[below=-1pt,pos=0.55]{京} (q3);
  \draw[->,thick] (q3) -- node[above=-1pt]{都} node[below=2pt,stp]{3} (q4);
  \draw[->,dashed,thick] (q2) -- node[fill=white,inner sep=1pt,font=\scriptsize,pos=0.38]{fail} (q3);

  \node[anchor=west,font=\scriptsize] at (3.45,0)    {accept 東京};
  \node[anchor=west,font=\scriptsize] at (3.45,-1.1) {accept 京都};

\end{tikzpicture}

\vspace{2mm}

%% ========= (b) precomputed table + per-frame flow =========
\begin{tikzpicture}[
  >={Stealth[round]}, font=\small,
  bx/.style={draw,rounded corners=1pt,inner sep=2pt,font=\scriptsize}]

  \node[font=\footnotesize\bfseries] at (-1.15,0.0) {(b)};

  % flow row (y = 0)
  \node                            (lg)   at (0.00,0) {$\ell_t$};
  \node[circle,draw,inner sep=1pt] (plus) at (0.98,0) {$+$};
  \node[bx]                        (am)   at (2.20,0) {$\arg\max$};
  \node                            (vt)   at (3.20,0) {$v_t$};
  \draw[->] (lg)   -- (plus);
  \draw[->] (plus) -- (am);
  \draw[->] (am)   -- (vt);

  % single precomputed table, cells = (bias, next state)
  \node[bx] (tab) at (1.78,-1.35) {%
    \begin{tabular}{c|cc}
      $q$   & 東                & 京都                        \\\hline
      $q_0$ & $2.00,\;q_1$      & $4.69{+}c,\;q_4$            \\
      $q_1$ & $0,\;q_1$         & $2.69{+}2c,\;q_4$           \\
    \end{tabular}};

  \draw[->] (tab.north -| plus) -- node[left,font=\scriptsize,inner sep=1pt]{$b(q_t,\cdot)$} (plus.south);
  \draw[->] (vt.south) -- node[right,font=\scriptsize,inner sep=1pt]{$q_{t+1}$} (tab.north -| vt);
\end{tikzpicture}

\caption{(a) Character-level Aho-Corasick automaton for 東京 and 京都; solid arrows are trie edges, the dashed one a failure link. Circled numbers trace the decoded stream 東京都 along the thick path, the failure link moving from the end of 東京 into the middle of 京都 without consuming a character, so both are matched. (b) The precomputed token-level table and per-frame decode flow; cells give the state reached and the bias $\psi(d')-\psi(d)$ at $w{=}1$, plus a lump $c$ per hotword completed.
}
\label{fig:automaton}
\end{figure}

Let $d_t=d(q_t)$ be the depth of the current state, $S=\{\emptyset, v_{t-1}\}$ the non-emitting set (blank or repeat token), $\hat{v}_t = \arg\max_v \ell_t(v)$ the unbiased proposal, and $\tilde{b}_t(v) = b_t(v)\mathbf{1}[v \in \text{top-}K(\ell_t)]$ the masked bias. The threshold $d^\star$ splits matches into \textit{entry} and \textit{deep} bands, and the gate assigns each mechanism to its own, replacing the argmax of Fig.~\ref{fig:automaton}b:

% Let $d_t = d(q_t)$ be the match depth at frame $t$ (distance from $q_0$ in Fig.~\ref{fig:automaton}a), $b_t(v) = b(q_t,v)$ the bias row for the current state, $S = \{\emptyset,\, v_{t-1}\}$ the non-emitting set (blank or a repeat of the previous token), and $\hat{v}_t = \arg\max_v \ell_t(v)$ the unbiased proposal. The masked bias is $\tilde{b}_t(v) = b_t(v)\,\mathbf{1}\!\left[v \in \text{top-}K(\ell_t)\right]$. The depth-adaptive gate assigns each mechanism to its band, replacing the single argmax of Fig.~\ref{fig:automaton}b:

\begin{equation*}
v_t =
\begin{cases}
\arg\max_v \left[\ell_t(v) + \tilde{b}_t(v)\right], & d_t < d^\star\\[3pt]
\hat{v}_t, & d_t \ge d^\star,\ \hat{v}_t \in S\\[3pt]
\arg\max_{v \notin S} \left[\ell_t(v) + b_t(v)\right], & \text{otherwise}
\end{cases}
\end{equation*}

This gives entry/insertion control below $d^\star$ and traversal protection above it. Setting $d^\star$=$\infty$ recovers masked biasing and $d^\star$=0 recovers two-stage, so $d^\star$ parameterizes the family containing both. Counting depth in raw characters would leave $d^\star$ script-dependent: a Han character carries $\approx$$1.75\times$ a kana's unigram surprisal on the benchmark transcripts, so we weight Han 1.0, kana 0.5 (unrounded 0.57 gives identical decodes) and threshold in Han-equivalent units. This is identical to character counting on uniform-script lists (ZH, verified byte-identical); on mixed scripts it replaces the separate kanji and kana optima with a single threshold $d^\star$=3 on the JA benchmark (§4.3).

%given a Han character carries about 1.75$\times$ the unigram surprisal of a kana on the benchmark transcripts (11.4 vs 6.5 bits), we weight Han 1.0 and kana 0.5, (unrounded 0.57 gives identical decodes)

\subsection{Reading-space biasing}
A calibrated surface push cannot help when the error is orthographic. The hypothesis carries the right reading in the wrong characters, and the surface automaton never engages (紹$\neq$照 at entry). We therefore run a second automaton in reading space. Offline, each hotword becomes a kana (MeCab/UniDic\footnote{https://taku910.github.io/mecab/}) or pinyin (tone-less pypinyin\footnote{https://github.com/mozillazg/python-pinyin}) path; each vocabulary token is assigned its candidate readings; and an alignment maps each surface character to its reading span. At decode time the automaton advances on any homophone token, anchoring on what the audio supports, while bias is spent only on the listed surface's tokens, an asymmetric match/push applied under the same top-K gate. It requires no runtime lexicon or trained module, and adds one further row lookup, transition and register per frame. It is the decode-time counterpart of \cite{bruguier2019phoebe,chen2019joint}, and distinct from prompting \cite{peng24b_interspeech}.

Because any homophone advances a reading match, each word's push is damped by $1/\bigl(1 + \log_{10}(1+f)\bigr)$, with $f$ its reading's
frequency in an independent corpus (ReazonSpeech transcripts \cite{yin2023reazonspeech} for JA, Aishell-1 training transcripts \cite{bu2017aishell} for ZH, eval split excluded); the list's per-word weight expresses domain intent. Both automata can run in a single greedy loop with biases combined pre-argmax. They are correlated evidence, so summing double-counts shared spans while the max under-credits corroboration; we fuse the surface and reading credits $s$ and $r$ as $\max(s,r) + \lambda \min(s,r)$, where $\lambda$ discounts the weaker source's corroboration for that correlation, $\lambda$=0 giving pure max and $\lambda$=1 pure sum. Gate evidence is unified in entropy units, so §2.2's gate applies unchanged.

\section{Open JA biasing benchmark}

The only JA biasing evaluation we know of \cite{nakagome25_interspeech} mines its keyword lists by decoding the evaluation sets with the model under test, which is circular by construction, and does not release them; the recent large contextual benchmark \cite{wang2025contextasr} covers ZH and EN only. We therefore construct and release the first open JA contextual-biasing evaluation set, built model-free from JSUT basic5000 \cite{sonobe2017jsut} and Common Voice (CV) 26 ja \cite{ardila2020common}. Model-free construction means the set is not easier for any particular system, unlike error-mined lists. 

\begin{table}[htb]
\centering
\caption{JA biasing benchmark statistics. JSUT is fully human-verified; CV is partly human-reviewed, the unreviewed residual noise rate is measured by a seeded blind sample and reported in the dataset card.}
\label{tab:bench}
\footnotesize
\setlength{\tabcolsep}{4.5pt}
\begin{tabular}{lrrrr}
\toprule
corpus & utts & w/ target & hotwords & occurrences \\
\midrule
JSUT     & 5,000   & 422 (8.44\%)    & 477   & 489    \\
CV ja & 300,077 & 25,876 (8.62\%) & 4,357 & 27,108 \\
\bottomrule
\end{tabular}
\end{table}

Because Japanese is unsegmented, the first question is what counts as a word. We annotate lexical-unit spans with MeCab/UniDic and merge compounds only where the merged form is itself corpus-attested, which keeps words like 東京大学 whole without absorbing the phrase around it. Target spans are formed after proper noun and rare-word filtering, followed by native-speaker screening. They are matched as annotated spans rather than substring hits, so a listed word occurring inside a longer word never counts. Per-utterance lists at $N \in \{100, 500, 1000, 2000\}$ fill with distractor words mined from Wikipedia, checked so that no distractor appears in its own utterance's reference and sampled to each corpus's target length profile, which keeps surface statistics from separating targets from distractors. The two corpora are complementary. JSUT is small, read and fully verified, while Common Voice contributes nearly two orders of magnitude more occurrences from crowd-sourced speech at the cost of a partially screened tail. Targets are labeled by length (2–6+ characters) and script, the regime variables analyzed in §4.3, made into a design principle. We release the per-utterance lists, span annotations, verification suite and card, seed-deterministic; audio remains under the source corpora's licenses. A more complete treatment of the methodology is planned for a companion resource paper.

% Targets are annotated span surfaces rather than substring hits, so a listed word inside a longer word never counts. Per-utterance lists at $N \in \{100, 500, 1000, 2000\}$ draw distractors from a shared leak-checked Wikipedia-mined pool matched to each corpus's target profile. Targets are labeled by length (2–6+ characters) and script, the regime variables analyzed in §4.3, made into a design principle. We release the per-utterance lists, span annotations, verification suite and card, seed-deterministic; audio remains under the source corpora's licenses. A more complete treatment of the methodology is planned for a companion resource paper.

\section{Experiment and results}

\subsection{Experimental setup}
We use two frozen public checkpoints, SenseVoice-small \cite{an2024funaudiollm} with a single-character CJK vocabulary and OWSM-CTC \cite{peng2024owsm} with multi-character BPE pieces, decoded with greedy CTC throughout. On ZH we evaluate Aishell-1 \cite{bu2017aishell} NE under the SeACo protocol \cite{shi2024seaco}, tuned on dev and reported on test, and WenetSpeech contextual set \cite{zhang2022wenetspeech} released with CB-Conformer \cite{xu2023cb} (298 words, 22k utterances), used untuned as a transfer cell. On JA, we evaluate §3, with a configuration frozen on an internal development set, barring $d^\star$=3, chosen from the length analysis (§4.3). CV recall is exact over all target-bearing utterances, with CER and precision measured on a seeded 55.9k-utterance evaluation sample. We report CER and min-clamped occurrence recall, together with biased-word precision and excess insertions. The latter two are jointly necessary, because the characteristic failure mode is corruption of ordinary text rather than false hotword hits.

\subsection{JA and ZH context-biasing results}

\begin{table*}[tb]
\centering
\caption{Decode-only biasing on frozen SenseVoice-small: Aishell-1 NE test, recall over the full list and over the hard R1 subset (trained references, R1 recall \cite{shi2024seaco}: CLAS 64, SeACo 80, +ASF 87); its untuned transfer to WenetSpeech \cite{xu2023cb}; the JA benchmark (\S3; N=100, configs frozen on internal benchmark). $K{=}10$; P=biased-word precision. Reading and combined (adaptive+reading) rows are frequency-damped (\S2.3). }
\label{tab:main}
\footnotesize
\setlength{\tabcolsep}{3.2pt}
\begin{tabular}{l c ccc ccc c ccc ccc}
\toprule
& & \multicolumn{3}{c}{Aishell-1 NE} & \multicolumn{3}{c}{WenetSpeech} & & \multicolumn{3}{c}{JSUT} & \multicolumn{3}{c}{Common Voice ja} \\
\cmidrule(lr){3-5} \cmidrule(lr){6-8} \cmidrule(lr){10-12} \cmidrule(lr){13-15}
method & config & CER & Rec.\ all/R1 & P & CER & Rec. & P & config & CER & Rec. & P & CER & Rec. & P \\
\midrule
greedy      & --                       & 7.54 & 52.4/8.9 & 99.6 & 5.62 & 88.6 & 99.7 & --                       & 8.42 & 38.0 & 100 & 18.50 & 29.1 & 99.9 \\
+ masked    & $w{=}2.5$                & 5.70 & 77.4/53.1 & 98.8 & 5.51 & 93.9 & 98.8 & $w{=}2$                  & 8.52 & 59.7 & 99.3 & 17.77 & \textbf{53.9} & 99.7 \\
+ two-stage & $w{=}4$                  & 5.91 & \textbf{83.5/66.5} & 96.9 & 6.43 & \textbf{94.5} & 93.5 & $w{=}2$                  & \textbf{8.47} & 54.6 & 99.6 & \textbf{17.62} & 50.5 & 99.8 \\
+ adaptive  & $w{=}3$, $d^\star{=}2$   & 5.48 & 81.0/60.6 & 98.7 & 5.69 & 94.1 & 98.3 & $w{=}2$, $d^\star{=}3$   & 8.52 & 60.3 & 99.3 & 17.69 & 53.5 & 99.7 \\
+ reading   & $w{=}2$                  & \textbf{5.20} & 79.0/58.1 & 98.9 & \textbf{5.41} & 93.5 & 98.6 & $w{=}1.5$        & 8.68 & 57.5 & 98.0 & 18.41 & 42.2 & 99.5 \\
+ combined  & $\lambda{=}0$            & 5.28 & 81.8/61.2 & 98.3 & 5.77 & 94.1 & 98.2 & $\lambda{=}0$            & 8.80 & \textbf{63.4} & 97.3 & 18.06 & 53.7 & 99.4 \\
\bottomrule
\end{tabular}
\end{table*}

Table 2 reports all configurations on both languages. On Aishell-1 NE, the unbounded two-stage gate raises recall on the hard R1 subset from 8.9\% to 66.5\% with SenseVoice, surpassing the 64\% of trained CLAS, without parameter updates. Every biased configuration also reduces CER relative to its own baseline (7.54\% → 5.20–5.91\%). The push resolves decisions the model had nearly made, rather than promoting words the acoustics did not support, which would raise CER. Precision is 96.9–98.9\% on ZH and 97.3–99.8\% on JA, and the masked configuration adds only 2 and 46 excess hotword emissions across the 5,000 and 55,876 utterances of JSUT and CV, against 0 and 5 unbiased.

We find, however, that no configuration is best everywhere, and hotword list composition plays a significant role in which is most effective. While two-stage leads on Aishell, whose targets are rare named entities, it gives up precision and CER on WenetSpeech, where the hotwords are common entities already recovered at 88.6\%. The push lands more often on words that need no help and converts into insertions, while top-K masking acts as a safety mechanism, improving on unbiased CER (5.51\%). On the JA benchmark, whose targets are mixed in script and length, the combined configuration leads or ties (63.4\% and 53.7\% recall) at a CER below unbiased decoding on CV (18.06\% vs 18.50\%). Splitting CV by whether an utterance contains a listed word localises that gain, with CER falling from 24.13\% to 22.34\% on target-bearing utterances and rising modestly from 14.15\% to 14.75\% on the rest. Composition thus acts twice. Rarity sets how hard a push may safely be, and heterogeneity how many kinds of evidence are needed, since targets spanning scripts and lengths fail for different reasons (§4.3).

The methods also transfer to the secondary OWSM model, whose vocabulary is multi-character rather than per-character. Applied unchanged, the masked configuration raises recall from 40.3\% to 53.4\% on JSUT and from 41.1\% to 62.4\% on CV at P$\geq$93.8\%, on baselines already above SenseVoice's (38.0\% and 29.1\%). The two-stage gate, by contrast, falls below unbiased decoding in both recall and CER. Which mechanism should act, and how strongly, therefore depends on evidence available at decode time, the question §2.2 poses and the next section characterizes.

\subsection{Regime analysis}

\begin{figure}[htbp]
\centering
\includegraphics[width=0.98\columnwidth]{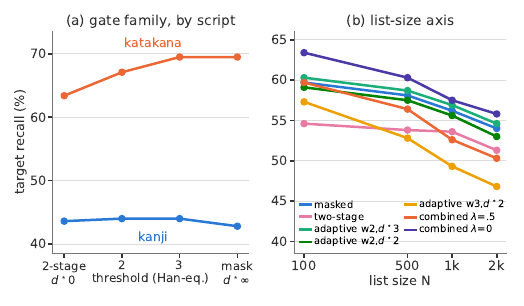}
\caption{Gate-family regimes. (a)~Threshold sweep at matched weight ($w{=}2$, $N{=}1000$); each endpoint degrades in the band the other protects. (b)~Effect of distractor list-size scaling on recall.}
\label{fig:regimes}
\end{figure}

Fig. 2a sweeps the gate threshold at matched weight between the two published endpoints, each of which degrades in the band the other protects. At $d^\star$=0 (two-stage) the entry band is unprotected and katakana recall falls from 69.5\% to 63.4\%; whereas at $d^\star$=$\infty$ (masked) the deep band is capped and kanji recall falls from 44.0\% to 42.8\%. Script determines which band a target occupies, and the benchmark stratifies by it; two-character targets accumulate too little bias for the gate to matter, and all surface configurations converge on them. An intermediate threshold recovers both maxima at once. At $d^\star$=3 the adaptive gate matches the masked endpoint on katakana (69.5\%) and exceeds both on kanji (44.0\% against 43.6\% and 42.8\%), collapsing in neither band.

The effect of distractors is evident as lists scale (Fig. 2b). At most a few of the N listed words occur in any utterance, so as N grows the model more often emits a character opening some listed word that was not spoken, and the automaton spends more of each utterance in spurious partial matches. Such a match pushes toward a wrong continuation, and completes into an insertion, so recall decays everywhere and precision decays faster. How fast depends on push magnitude rather than on the gate. Over a twentyfold increase in N, adaptive at $w$=3 loses 10.5 points of recall and falls to 80\% precision, while every $w$=2 configuration caps losses at 3 to 6 points, and holds precision above 95\%. A smaller $w$ cuts spurious overwrites before rescues, since genuine corrections need only a small push. Delaying the threshold cannot substitute, moving recall by 1 to 3 points where the weight moves 6 to 10, which corroborates the per-scale weight re-tuning in \cite{andrusenko2025turbobias}. The same sensitivity appears across architectures, where OWSM's multi-character pieces deliver several characters of push in one emission and corrupt the hypothesis outright, inflating it to 1.88 times the reference length.

\subsection{Reading-space matching and combination}
Among occurrences the surface decode misses, the proportion whose reading is nevertheless present is 72\% on NE-dense Aishell and 31\% and 18\% on JSUT and CV. The gap separates two kinds of miss. Rare proper nouns, which dominate the JA sets, are misheard outright, whereas orthographic variants are heard correctly but misspelled, the model writing 理屈 for the listed 理窟 and 肩透かし for 肩すかし. Reading-space matching hence addresses spelling rather than recognition errors, which is why it trails surface matching on these sets and why its distinct contribution lies in that second class. For 27 listed words on Aishell, 20 on JSUT and 88 on CV, surface biasing recovers no occurrence, while reading-space matching recovers 31 of 32, all 22, and 131 of 353 respectively. Its gains concentrate on short kanji targets, the condition where all surface gates tie, and are absent on long katakana names, where reading and surface channels coincide.

% Among occurrences the surface decode misses, the proportion whose reading is nevertheless present is 72\% on NE-dense Aishell and 31\% and 18\% on JSUT and CV. The gap separates two kinds of miss. For orthographic variants, like the listed 理窟 which the model writes as the standard 理屈, reading-space matching recovers 14 of 20 occurrences whereas surface biasing recovers none. The rare proper nouns that dominate the JA sets are instead misheard outright. Reading-space matching hence addresses spelling rather than recognition errors, which is why it trails surface matching there. Its gains concentrate on short kanji targets, the condition where all surface gates tie, and are absent on long katakana names, where reading and surface channels coincide. 

Because they miss different occurrences, combining surface and reading channels matches or improves on either alone across all datasets. Whenever both channels push the same token, $\lambda$ adds a share of the weaker score, treating agreement as evidence. As N grows both automata are usually inside some partial match, so a token advancing both signals coincidence more often than agreement. Taking the maximum ($\lambda$=0) ignores the weaker channel altogether, and matches or exceeds $\lambda$>0 at every list size on both corpora while restoring precision (63.4\% at 97.3\% against 59.7\% at 95.5\% on JSUT). Loosening the match itself degrades results; admitting fuzzy pinyin at equal credit was found to lower every metric, since fuzzy transitions move the automaton off the true path as readily as onto it. On WenetSpeech, frequency damping cuts excess insertions from 714 to 304 and raises precision from 96.9\% to 98.6\% at a cost of 0.2 points of recall; on the rare-entity Aishell list it cuts them from 19 to 8, target rarity having already kept the count low.

\section{Conclusion}
We present boundary-free contextual biasing for frozen CTC models, where a character automaton exposes match depth, unifying two gating mechanisms that suit different regimes, and a per-mora or syllable reading state that recovers homophone errors. Combining the two never loses recall against the gate alone and adds up to 3 points. It reaches 66.5\% R1 recall on Aishell-1 NE, above a trained baseline, transfers to WenetSpeech and OWSM unretuned, and gains 25 points on the JA benchmark we release. List composition determines which configuration wins, gains decay as lists grow because push magnitude is a tuned constant, and multi-character vocabularies still need the masked endpoint. Both mechanisms assume dense scripts. In a language like Thai where a character carries less evidence, partial matches are common and depth accrues credit too fast, while a phonemic orthography leaves little for reading-space matching; the schedule would need recalibrating in information units.

\vfill\pagebreak

% \section{REFERENCES}
% \label{sec:refs}

% List and number all bibliographical references at the end of the
% paper. The references can be numbered in alphabetic order or in
% order of appearance in the document. When referring to them in
% the text, type the corresponding reference number in square
% brackets as shown at the end of this sentence \cite{C2}. An
% additional final page (the fifth page, in most cases) is
% allowed, but must contain only references to the prior
% literature.

\bibliographystyle{IEEEbib}
\bibliography{refs}

@article{aho-corasick,
author = {Aho, Alfred and Corasick, Margaret},
year = {1975},
pages = {333-340},
title = {Efficient string matching: An aid to bibliographic search},
volume = {18},
journal = {Commun. ACM},
doi = {10.1145/360825.360855}
}

@inproceedings{ctc,
author = {Graves, Alex and Fernández, Santiago and Gomez, Faustino and Schmidhuber, Jürgen},
year = {2006},
pages = {369-376},
title = {Connectionist {T}emporal {C}lassification: Labelling unsegmented sequence data with recurrent neural networks},
booktitle = {Proceedings of the 23rd International Conference on Machine Learning (ICML)},
doi = {10.1145/1143844.1143891}
}

@inproceedings{aleksic15_interspeech,
  title     = {{Bringing contextual information to Google speech recognition}},
  author    = {Petar Aleksic and Mohammadreza Ghodsi and Assaf Michaely and Cyril Allauzen and Keith Hall and Brian Roark and David Rybach and Pedro Moreno},
  year      = {2015},
  booktitle = {{Interspeech}},
  pages     = {468--472},
  doi       = {10.21437/Interspeech.2015-177},
  issn      = {2958-1796},
}

@inproceedings{zhao19d_interspeech,
  title     = {{Shallow-fusion end-to-end contextual biasing}},
  author    = {Ding Zhao and Tara N. Sainath and David Rybach and Pat Rondon and Deepti Bhatia and Bo Li and Ruoming Pang},
  year      = {2019},
  booktitle = {{Interspeech}},
  pages     = {1418--1422},
  doi       = {10.21437/Interspeech.2019-1209},
  issn      = {2958-1796},
}

@inproceedings{pundak2018deep,
  title={Deep context: End-to-end contextual speech recognition},
  author={Pundak, Golan and Sainath, Tara N and Prabhavalkar, Rohit and Kannan, Anjuli and Zhao, Ding},
  booktitle={IEEE Spoken Language Technology Workshop (SLT)},
  pages={418--425},
  year={2018},
  organization={IEEE}
}

@inproceedings{bruguier2019phoebe,
  title={Phoebe: Pronunciation-aware contextualization for end-to-end speech recognition},
  author={Bruguier, Antoine and Prabhavalkar, Rohit and Pundak, Golan and Sainath, Tara N},
  booktitle={IEEE International Conference on Acoustics, Speech and Signal Processing (ICASSP)},
  pages={6171--6175},
  year={2019},
  organization={IEEE}
}

@inproceedings{chen2019joint,
  title={Joint grapheme and phoneme embeddings for contextual end-to-end asr},
  author={Chen, Zhehuai and Jain, Mahaveer and Wang, Yongqiang and Seltzer, Michael L and Fuegen, Christian},
  booktitle={Interspeech},
  pages={3490--3494},
  year={2019}
}

@inproceedings{sun2021tree,
  title={Tree-constrained pointer generator for end-to-end contextual speech recognition},
  author={Sun, Guangzhi and Zhang, Chao and Woodland, Philip C},
  booktitle={IEEE Automatic Speech Recognition and Understanding Workshop (ASRU)},
  pages={780--787},
  year={2021},
  organization={IEEE}
}

@inproceedings{context_trie,
author = {Le, Duc and Jain, Mahaveer and Keren, Gil and Kim, Suyoun and Shi, Yangyang and Mahadeokar, Jay and Chan, Julian and Shangguan, Yuan and Fuegen, Christian and Kalinli, Ozlem and Saraf, Yatharth and Seltzer, Michael},
year = {2021},
pages = {1772-1776},
title = {Contextualized streaming end-to-end speech recognition with trie-based deep biasing and shallow fusion},
doi = {10.21437/Interspeech.2021-1566},
booktitle = {{Interspeech}}
}

@inproceedings{xu2023cb,
  title={{CB}-Conformer: Contextual biasing conformer for biased word recognition},
  author={Xu, Yaoxun and Liu, Baiji and Huang, Qiaochu and Song, Xingchen and Wu, Zhiyong and Kang, Shiyin and Meng, Helen},
  booktitle={IEEE International Conference on Acoustics, Speech and Signal Processing (ICASSP)},
  pages={1--5},
  year={2023},
  organization={IEEE}
}

@inproceedings{shi2024seaco,
  title={{SeACo-Paraformer}: A non-autoregressive asr system with flexible and effective hotword customization ability},
  author={Shi, Xian and Yang, Yexin and Li, Zerui and Chen, Yanni and Gao, Zhifu and Zhang, Shiliang},
  booktitle={IEEE International Conference on Acoustics, Speech and Signal Processing (ICASSP)},
  pages={10346--10350},
  year={2024},
  organization={IEEE}
}

@inproceedings{andrusenko2025turbobias,
  title={Turbobias: Universal asr context-biasing powered by gpu-accelerated phrase-boosting tree},
  author={Andrusenko, Andrei and Bataev, Vladimir and Grigoryan, Lilit and Lavrukhin, Vitaly and Ginsburg, Boris},
  booktitle={IEEE Automatic Speech Recognition and Understanding Workshop (ASRU)},
  pages={1--7},
  year={2025},
  organization={IEEE}
}

@article{an2024funaudiollm,
  title={Funaudio{LLM}: Voice understanding and generation foundation models for natural interaction between humans and llms},
  author={An, Keyu and Chen, Qian and Deng, Chong and Du, Zhihao and Gao, Changfeng and Gao, Zhifu and Gu, Yue and He, Ting and Hu, Hangrui and Hu, Kai and others},
  journal={arXiv preprint arXiv:2407.04051},
  year={2024}
}

@inproceedings{peng24b_interspeech,
  title     = {{OWSM v3.1: Better and Faster Open Whisper-Style Speech Models based on E-Branchformer}},
  author    = {Yifan Peng and Jinchuan Tian and William Chen and Siddhant Arora and Brian Yan and Yui Sudo and Muhammad Shakeel and Kwanghee Choi and Jiatong Shi and Xuankai Chang and Jee-weon Jung and Shinji Watanabe},
  year      = {2024},
  booktitle = {{Interspeech}},
  pages     = {352--356},
  doi       = {10.21437/Interspeech.2024-1194},
  issn      = {2958-1796},
}

@inproceedings{bu2017aishell,
  title={Aishell-1: An open-source mandarin speech corpus and a speech recognition baseline},
  author={Bu, Hui and Du, Jiayu and Na, Xingyu and Wu, Bengu and Zheng, Hao},
  booktitle={O-COCOSDA},
  pages={1--5},
  year={2017},
  organization={IEEE}
}

@inproceedings{zhang2022wenetspeech,
  title={Wenetspeech: A 10000+ hours multi-domain mandarin corpus for speech recognition},
  author={Zhang, Binbin and Lv, Hang and Guo, Pengcheng and Shao, Qijie and Yang, Chao and Xie, Lei and Xu, Xin and Bu, Hui and Chen, Xiaoyu and Zeng, Chenchen and others},
  booktitle={IEEE International Conference on Acoustics, Speech and Signal Processing (ICASSP)},
  pages={6182--6186},
  year={2022},
  organization={IEEE}
}

@article{sonobe2017jsut,
  title={{JSUT} corpus: free large-scale Japanese speech corpus for end-to-end speech synthesis},
  author={Sonobe, Ryosuke and Takamichi, Shinnosuke and Saruwatari, Hiroshi},
  journal={arXiv preprint arXiv:1711.00354},
  year={2017}
}

@inproceedings{ardila2020common,
  title={Common Voice: A massively-multilingual speech corpus},
  author={Ardila, Rosana and Branson, Megan and Davis, Kelly and Kohler, Michael and Meyer, Josh and Henretty, Michael and Morais, Reuben and Saunders, Lindsay and Tyers, Francis and Weber, Gregor},
  booktitle={Proceedings of the 12th Language Resources and Evaluation Conference (LREC)},
  pages={4218--4222},
  year={2020}
}

@inproceedings{yin2023reazonspeech,
  author    = {Yue Yin and Daijiro Mori and Seiji Fujimoto},
  title     = {ReazonSpeech: A free and massive corpus for Japanese ASR},
  booktitle = {Proceedings of the 29th Annual Meeting of the Association for Natural Language Processing},
  year      = {2023},
  url       = {https://research.reazon.jp/_static/reazonspeech_nlp2023.pdf}
}

@inproceedings{nakagome25_interspeech,
  title     = {{WCTC-Biasing: Retraining-free Contextual Biasing ASR with Wildcard CTC-based Keyword Spotting and Inter-layer Biasing}},
  author    = {Yu Nakagome and Michael Hentschel},
  year      = {2025},
  booktitle = {{Interspeech}},
  pages     = {5178--5182},
  doi       = {10.21437/Interspeech.2025-1300},
  issn      = {2958-1796},
}

@article{wang2025contextasr,
  title={Context{ASR}-Bench: A massive contextual speech recognition benchmark},
  author={Wang, He and Ma, Linhan and Guo, Dake and Wang, Xiong and Xie, Lei and Xu, Jin and Lin, Junyang},
  journal={arXiv preprint arXiv:2507.05727},
  year={2025}
}

@inproceedings{peng2024owsm,
  title={{OWSM-CTC}: An open encoder-only speech foundation model for speech recognition, translation, and language identification},
  author={Peng, Yifan and Sudo, Yui and Shakeel, Muhammad and Watanabe, Shinji},
  booktitle={62nd Annual Meeting of the Association for Computational Linguistics (ACL)},
  pages={10192--10209},
  year={2024}
}

\end{document}